\documentstyle[aps,twocolumn,psfig]{revtex}

\begin{document}
\title{Excitonic photoluminescence in symmetric coupled
double quantum wells subject to an external electric field}

\draft
\author{J. Soubusta\cite{soub},
R. Grill, P. Hl\'{\i}dek, M. Zv\'{a}ra}
\address{\it Institute of Physics, Charles University, Ke Karlovu 5,
121~16 Prague 2, Czech Republic}
\author{L. Smr\v{c}ka}
\address{\it Institute of Physics, Academy of Science of the Czech
Republic, Cukrovarnick\'{a} 10, 162~53~Prague~6, Czech Republic}
\author{S. Malzer, W. Gei{\ss}elbrecht, G. H. D\"{o}hler}
\address{Institut f\"{u}r Technische Physik I, Universit\"{a}t
Erlangen-N\"{u}rnberg, Erwin-Rommel-Stra{\ss}e 1, 91058~Erlangen,
Germany}
\date{\today}

\maketitle

\begin{abstract}
The effect of an external electric field $F$ on the excitonic
photoluminescence (PL) spectra of a symmetric coupled double
quantum well (DQW) is investigated both theoretically and
experimentally. We show that the variational method in a two-particle
electron-hole wave function approximation gives a good agreement
with measurements of PL on a narrow DQW in a wide interval of
$F$ including flat-band regime. The experimental data are presented
for an MBE-grown DQW consisting of two 5~nm wide GaAs wells,
separated by a 4 monolayers (MLs) wide pure AlAs central barrier,
and sandwiched between Ga$_{0.7}$Al$_{0.3}$As layers. The bias
voltage is applied along the growth direction. Spatially direct and
indirect excitonic transitions are identified, and the radius of
the exciton and squeezing of the exciton in the growth direction
are evaluated variationally. The excitonic binding energies,
recombination energies, oscillator strengths, and relative
intensities of the transitions as functions of the applied field
are calculated. Our analysis
demonstrates that this simple model is applicable in case of narrow
DQWs not just for a qualitative description of the PL peak
positions but also for the estimation of their individual shapes and
intensities.

\end{abstract}
\pacs{PACS numbers: 73.20.Dx, 78.55.-m, 78.66.-w, 78.66.Fd}

\section{Introduction}

Because of their technical importance and unique physical
properties, semiconductor quantum wells have been the subject of
intensive research since their first fabrication in 1973.
A brief review of the main achievements and a number of representative
references can be found, e.g., in books \cite{bast} and \cite{weis}.
The tunnel-coupled quantum states in DQWs are very sensitive to both
electric and magnetic fields, and changes induced by these fields have
been intensively studied by PL, PL excitation, and photoconductivity
\cite{lee,dig,west,taka,bay,akim}.

A symmetric coupled DQW consists of two identical quantum wells
separated by only a thin barrier. The energy levels of the coupled
QWs split owing to the interwell tunneling. In the flat-band
condition $(F=0)$, the eigenfunctions of the DQW have defined symmetries.
In this situation, only transitions between electron and hole states
of the same symmetry are optically allowed. The ground-state wave
function is symmetric, while that of the first excited state is
antisymmetric. When an electric field is applied to the DQW,
the wave functions become predominantly localized in one well
only. Transitions allowed in the flat-band case evolve into
spatially indirect (interwell) transitions, which shift linearly in
energy as a function of the applied field. On the other hand,
transitions forbidden in the flat-band case become allowed and
represent spatially direct (intrawell) transitions. Two factors
determine the optical transition intensities: (a) the overlap
integral of the electron and hole single-particle wave functions,
and (b) the population of the energy subbands for a given
temperature. In order to describe the optical transitions correctly
one has to account for the Coulomb interaction between electrons
and holes. We use a two-particle wave function composed of the
electron and hole single-particle wave functions multiplied by a
function of their relative positions to get an exciton binding
energy.

The purpose of this report is to demonstrate that our quite simple
theory, applied to the PL results obtained from our specifically
designed sample, provides excellent quantitative agreement with
regard to both the energetic position of the PL peaks and their
shapes and intensities as a function of applied electric fields.
These results represent clearly a further improvement compared to
the previously reported results \cite{lee,dig,taka,akim}.

\section{Theory}

The calculation starts from the envelope-function excitonic
Hamiltonian
\begin{equation}
H_{ex} = H_{oe} +H_{oh} +H_{2D} +U +E_g,
\label{Hex}
\end{equation}
where the respective electron and hole single-particle terms read
\cite{bast}
\begin{equation}
H_{o\nu}=-\frac{\hbar^2}2 \frac{\partial}{\partial z_\nu}
       \frac1{m_\nu(z_\nu)} \frac{\partial}{\partial z_\nu}
       +V_\nu(z_\nu) -q_\nu Fz_\nu ,
\label{H0}
\end{equation}
$\nu =\{ e,h\}$ denotes an electron or hole, and $q_\nu$ is the
respective charge. $m_\nu(z_\nu)$ and $V_\nu(z_\nu)$ are the respective
electron or hole $z_\nu$ dependent effective masses and confining
potentials of the DQW structure. The electric field $F$ is applied
along the $z$ axis parallel to the growth direction. $H_{2D}$
represents the kinetic energy in the $xy$ plane. The electron-hole
exciton interaction is included by means of the Coulomb term $U$,
and the GaAs bandgap energy $E_g$ completes the total energy
$H_{ex}$.

\begin{figure}[h]
\psfig{file=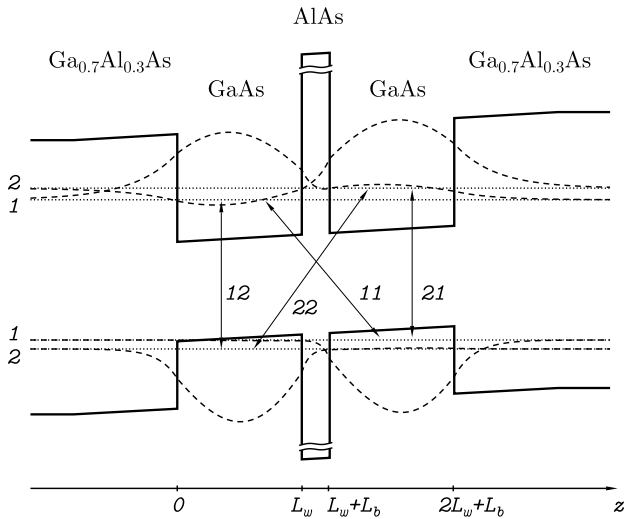,width=8.4cm,angle=270,clip=}
\caption{Scheme of the DQW potential profile with an applied
external field (solid lines). The energy levels (dotted lines) and
wave functions (dashed lines) of electrons and holes are shown as well.
Bases of the wave functions are at the corresponding energies and their
amplitudes are plotted in arbitrary units. The plot was calculated for
$L_w=5$~nm and $F=30$~kV/cm.}
\label{bands}
\end{figure}

Eigenenergies and eigenfunctions of the single-particle 1D
Hamiltonian $H_{o\nu}$ are found by using linear combinations of analytical
functions sin$(\xi)$ and cos$(\xi)$ (wells, $F=0$), Airy functions
Ai$(\xi)$ and Bi$(\xi)$ (wells, $F>0$), or exp$(\pm\xi)$
(barriers), and by matching the wave function amplitudes and their
derivatives divided by the effective masses at each interface.
Representative results of such a calculation made for a DQW
structure with $L_w=18$~ML, $L_b=4$~ML ($\approx$ 5.09~nm, and
1.13~nm, respectively), and for $F=30$~kV/cm are schematically
plotted in Fig.~\ref{bands}.

The first two single-particle eigenfunctions $\varphi_1^\nu(z_\nu)$,
$\varphi_2^\nu(z_\nu)$ of $H_{o\nu}$ are used to construct a set
of basis functions for the variational calculation of the exciton
states as a single product of one electron and one hole wave function
multiplied by a function of relative electron hole positions
\cite{dig,west,taka}:
\begin{equation}
\chi_{ij}=N_{ij}\varphi_i^e(z_e) \varphi_j^h(z_h) \exp
\left(-\frac{\sqrt{\rho^2+\alpha_{ij}(z_e-z_h)^2}}{R_{ij}}\right),
\label{chi}
\end{equation}
where the normalization factors $N_{ij}$ and the variational
parameters $R_{ij}, \alpha_{ij}$ are, in general, different for
each exciton. While the $z$ coordinates of both carriers are
specified absolutely $(z_e,z_h)$, only the relative distance
between them projected onto the $xy$ plane $(\rho )$ is relevant.
We minimize the total energy [Eq. (\ref{Hex})] for each
exciton separately and obtain four wave functions
\{$\chi_{11},\chi_{12},\chi_{21},\chi_{22}$\}.

In our calculation, we neglect the heavy- and light-hole bands
mixing and assume strictly parabolic dispersion relations. Also,
the mixing of the exciton states by the Coulomb interaction is not
taken into account because of its small effect on the narrow DQW
system\cite{west}. The GaAs/AlGaAs/AlAs material parameters and
their compositional dependence is taken as in
\cite{bast,west,bajaj,gap}.

The $\chi_{ij}$s are used to evaluate the optical oscillator
strength (see, e.g., Ref.~\cite{dig}) and the electron-hole overlap
integral $F_{ij}(0)$, which is given by
\begin{equation}
F_{ij}(0) = \!\!\int\limits_{-\infty}^{\infty}\!\!\!
            \chi_{ij} \mbox{(z$_e$=z,z$_h$=z,$\rho$=0)} dz
          = N_{ij}\!\!\int\limits_{-\infty}^{\infty}\!\!\!
            \varphi_i^e(z)\varphi_j^h(z) dz.
\label{overlap}
\end{equation}
The normalized exciton PL intensity spectrum is then calculated as
\begin{eqnarray}
I(E) &=& \frac{\sum\limits_{i,j=1}^2
         L_{ij}(E) |F_{ij}(0)|^2 e^{-\beta E_{ij}}}
         {\sum\limits_{i,j=1}^2 |F_{ij}(0)|^2 e^{-\beta E_{ij}}},\nonumber\\
L_{ij}(E) &=& \frac {\beta\Delta_{ij}}{\pi} \int\limits_0^{\infty}
              \frac {e^{-\beta E'}dE'}{(E-E_{ij}-E')^2+\Delta_{ij}^2},
\label{spec}
\end{eqnarray}
$\beta=1/k_BT$, $k_B$ is the Boltzmann constant and $T$ the
temperature. The convolution form $L_{ij}(E)$ expresses the
normalized line-shape of the $i,j$th transition, which is determined
by a thermal distribution of the excitonic kinetic energy and slightly
diffused by a Lorenz function. The width $\Delta_{ij}=0.5$~meV
is used in every case.

\section{Experimental}

Our PL experiments were performed on a sample grown by MBE at a
temperature of 600$^o$C on a semi-insulating GaAs substrate
oriented in the [001] direction. The growth started with a 500~nm
wide n-doped (Si, $2\times 10^{18}$~cm$^{-3}$) GaAs layer, followed
by a 300~nm wide n-doped (Si, $1.4\times 10^{18}$~cm$^{-3}$) GaAlAs
layer. The Al content in the GaAlAs layers was always 0.3. After
this the following sequence of GaAlAs layers was grown: 500~nm
intrinsic, 5~nm p-$\delta$-doped (C, $3\times 10^{17}$~cm$^{-3}$)
layer, and 100~nm intrinsic. On top of this separating layer, a
sequence of three symmetric DQWs with 4~MLs ($\approx$ 1.13~nm)
wide pure AlAs central barriers in between were grown, employing
growth interruptions of 20~s at each heterointerface. The well
widths are 35~MLs, 26~MLs, and 18~MLs ($\approx$ 10~nm, 7.5~nm, and
5~nm), respectively. The DQWs are separated by a 100~nm wide GaAlAs
layer in each case. The growth then continued with another sequence
of GaAlAs layers: 100~nm intrinsic, 5~nm n-$\delta$-doped (Si,
$4\times 10^{17}$~cm$^{-3}$), and 500~nm intrinsic. On the top a
300~nm wide p-doped (C, $1.4\times 10^{18}$~cm$^{-3}$) GaAlAs layer
and a 20~nm wide p-doped (C,$2\times 10^{18}$~cm$^{-3}$) GaAs cap
layer were grown.

The {\it p-i-n} configuration of the sample allows us to apply a
bias voltage $U_{pn}$ by means of selective Ohmic contacts to the
p-doped layers on the top and to the n-doped layers at the bottom.
Usually, the bands in any {\it p-i-n} structure are tilted and
application of forward bias is required to flatten them. We,
therefore, used a special sample design comprising $\delta$-doped
layers inside the intrinsic region of the structure in order to
screen the built-in electric field. Thus, the flat-band regime is
obtained almost at $U_{pn}=0$~V, avoiding high dark current present
at higher forward bias. The measured devices of the size
$250\;\mu$m~$\times 250\;\mu$m were defined photolithographically
and mesa-isolated. Detailed results will only be presented for the
narrow DQW with $L_w=5$~nm and $L_b=1.13$~nm, for which the
application of our simple theoretical model is justified.

\begin{figure}[b]
\psfig{file=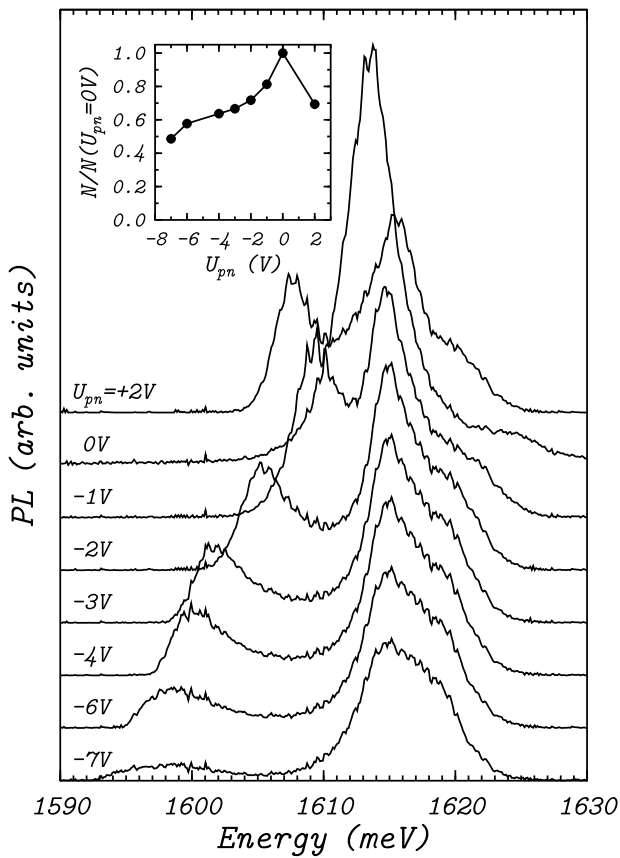,width=8.4cm,clip=}
\caption{PL spectra of the DQW with 5~nm wide wells as a function
of various bias voltages. The applied voltages $U_{pn}$ are denoted
beside each curve. The inset shows the normalized integral
intensity $N/N(U_{pn}=0V)$ of the presented PL curves.}
\label{exp1}
\end{figure}

Figure~\ref{exp1} shows the PL spectra of the sample for various
bias voltages. The sample was cooled in a closed-cycle cryostat and
excited by a Ti:sapphire laser pumped by an Ar$^+$-ion laser. We
used an excitation power of $\sim 100$~mW/cm$^{2}$ at a photon
energy of 1722~meV (below the bandgap energy of
Ga$_{0.7}$Al$_{0.3}$As). The emitted luminescence was analyzed by a
monochromator with 0.6~m focal length and a 1200~grooves/mm
grating, and detected by a cooled CCD camera. We applied reverse
bias up to $U_{pn}=-8$~V with the maximum current of only $\sim
10~\mu$A. An estimation of a dissipated power and a negligible
shift in position of the spectrum measured with $U_{pn}=+2$~V
proved that Ohmic heating of the sample for both reverse and
forward biased junction is insignificant.

\section{Discussion}

The sample was designed as a single (not multiple) DQW structure in
order to minimize the well width fluctuations. For the same reason,
the growth interruptions were employed, and a pure AlAs barrier
instead of an Al$_x$Ga$_{1-x}$As one was chosen to improve its
homogeneity and to avoid barrier height fluctuations. The
broadening of the direct exciton peak of $\sim 4$~meV (see
Fig.~\ref{exp1}) and even lower at 10~K ($\sim 2$~meV, not shown)
indicates a quite good quality of the sample. Furthermore, the
exciton peaks are not split due to large area monolayer
fluctuations of the well widths, a phenomenon which is frequently
observed on samples grown with interruptions.

The $\delta$-doped layers shielding the DQWs within the {\it p-n}
junction are effective in case of an excitation below the
Ga$_{0.7}$Al$_{0.3}$As bandgap energy, when the carriers are
excited exclusively in the wells. The flat band condition is found
almost at zero bias voltage. On the other hand, when using a He-Ne
laser excitation above bandgap energy, the $\delta$-doped layers
are neutralized and become inactive. Consequently, a forward bias
corresponding to the bandgap energy is needed in order to overcome
the built-in field and to establish the flat-band regime.

\begin{figure}[h]
\psfig{file=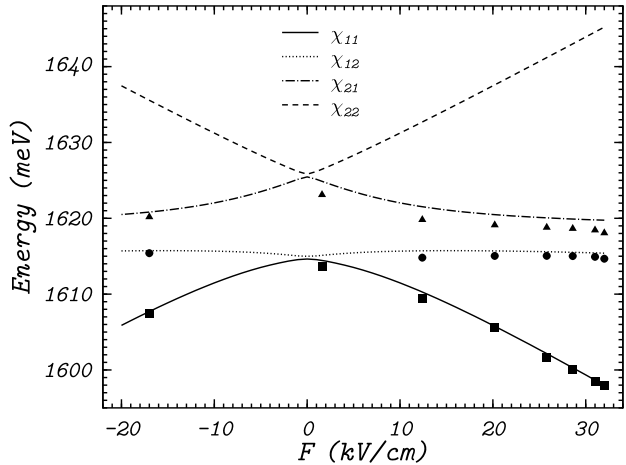,width=8.4cm,angle=270,clip=}
\caption{PL peak positions as a function of electric field $F$.
The lines are calculated for the DQW depicted in
Fig.~\protect\ref{bands} at $T=45$~K. The markers indicate
the peak positions of the experimental PL spectra plotted
in Fig.~\protect\ref{exp1}.}
\label{energ}
\end{figure}

The experimental spectra plotted in Fig.~\ref{exp1} were analyzed
to gain maximum data for the theoretical procedure. In the
calculation, we kept the parameters $L_w$ and $L_b$ constant and
varied only $T$ and $F$ for optimization. The temperature
$T=45$~K of the sample was obtained by a comparison of the
measured and calculated relative peak intensities of the spectra
in the linear field range. The field $F$ was deduced
by means of tracing the energy distance between the first two
exciton transitions, $\chi_{11}$ and $\chi_{12}$.
A very good agreement of the calculated and observed
transition energies is represented by Fig.~\ref{energ}. Notice
that the optimized $T$ provides an excellent correspondence in the
peak positions. The absolute positions of all the peaks including
the resonant splitting of the symmetric-antisymmetric states were
obtained without any free parameter. The small difference in the
intensities and positions of the theoretical lines $\sim 1$~meV
in comparison with the experiment can be attributed to (i) the
approximations used in the theory and (ii) small deviations from
the intended structure of the sample.

\begin{figure}[h]
\psfig{file=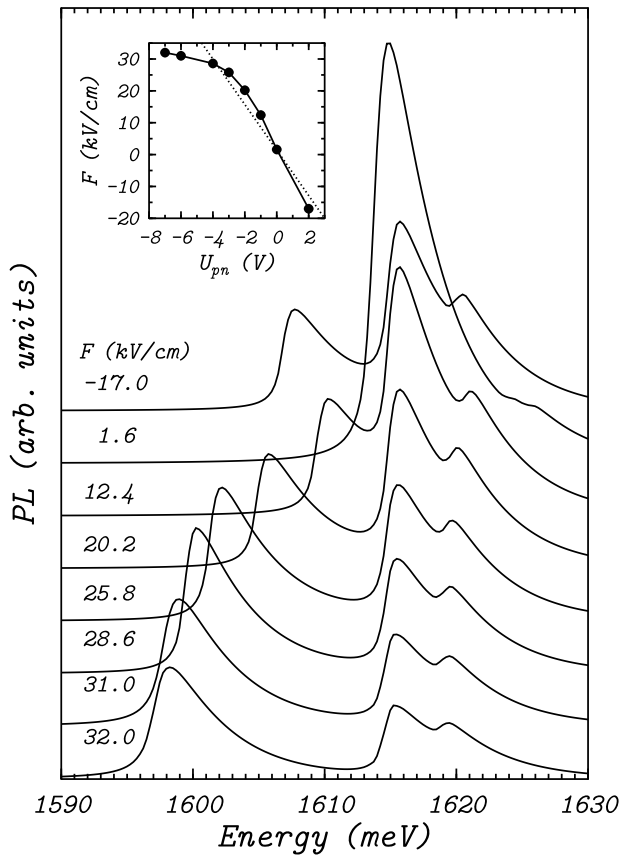,width=8.4cm,clip=}
\caption{Calculated PL spectra for the DQW depicted in Fig.~\protect
\ref{bands} at a temperature $T=45$~K. The applied electric field
$F$ is given beside each curve. The inset shows the electric field
induced in the DQW by the bias voltage $U_{pn}$.}
\label{teori}
\end{figure}

The inset of Fig.~\ref{teori} shows the deduced field $F$
as a function of the bias voltage applied to the sample. In the
case of the spectrum close to flat-band regime ($U_{pn}=0V$),
the corresponding field ($F=1.6$~kV/cm) was determined from a linear
part of this dependence. As anticipated, the dependence is linear
within a limited voltage interval ($-3V\leq U_{pn}\leq 2V$),
and saturates for higher reverse bias. The linear
regime of the band tilting is in very good agreement with $F$
calculated directly from the sample design (dotted line in the
inset of Fig.~\ref{teori}). A small discrepancy can be explained as
an effect of a non-homogeneous distribution of the electric field
over the three DQW systems, or as a result of a too simple
description of the single wave functions in the applied variational
method.

Figure~\ref{teori} shows the spectra calculated for the studied DQW
structure for the optimized temperature and the fields
corresponding to individual PL curves plotted in Fig.~\ref{exp1}.
The spectra are magnified in accordance with the measured integral
intensities $N(U_{pn})$ (see the inset of Fig.~\ref{exp1}) to allow
a comparison with the measured PL. The optimized temperature seems
to be reasonable taking into account that the sample in a
closed-cycle cryostat is cooled by a cold finger and placed on a
rather low thermally conductive substrate. The close concert of the
experimental and calculated PL spectra in the low field interval
($-3V\leq U_{pn}\leq 2V$) is evident. By increasing reverse bias
the observed indirect transition diminishes. This results from a
low stability of the lowest energy $\chi_{11}$ exciton which
becomes dissociated by electron tunneling outside the DQW. The
highest transition $\chi_{22}$ appears near the flat-band case only.
It manifests itself in Fig.~\ref{teori} in the spectrum calculated for
$F=1.6$~kV/cm as the small peak at the energy of 1626~meV close to the
$\chi_{21}$ transition. In the experiment, these transitions merge
into one peak observed at 1623~meV. Out of the flat-band case,
$\chi_{22}$ is suppressed due to a weak population of the higher
levels and the lower value of the electron-hole overlap integral
$F_{22}(0)$ [Eq. (\ref{overlap})]. With respect to the good agreement of
relative intensities of the spatially direct and indirect
transitions we can conclude that the overlap integrals are well
estimated and the particle distribution is described by our wave
functions in a satisfactory way.

\section{conclusion}

In this paper we have concentrated on the photoluminescence of
narrow symmetric coupled double quantum wells, where the energy
difference between excitonic states is high enough to allow
neglecting of band mixing effects. The inclusion of exciton
interaction is unavoidable for a correct description of the
electron-hole states and positioning of optical transitions. We
have shown that the simple model describes nearly perfectly all
dominant features of the optical transitions, namely relative
intensities, absolute positions of the individual transitions, and
the resonant splitting of the symmetric-antisymmetric states.

\acknowledgments
This work was supported by the Friedrich-Alexander-Universit\"{a}t
Erlangen-N\"{u}rnberg, by the Grant Agency of the Czech Republic (Grant
No. 202/95/1533 and No. 202/98/0085), and by the Ministry of
Education of the Czech Republic (Grant No. VS-97113). R. Grill
acknowledges support from the Alexander von Humboldt Foundation
(AvH, Bonn, Germany).


\begin{thebibliography}{99}

\bibitem[*]{soub} E-mail address: soubusta@alma.karlov.mff.cuni.cz

\bibitem{bast} G. Bastard, {\em Wave Mechanics Applied to
Semiconductor Heterostructures, Monographies de physique, Paris}
(1992).

\bibitem{weis} C. Weisbuch and B. Winter, {\em Quantum
Semiconductor Structures, Academic Press, New York} (1991).

\bibitem{lee} J. Lee, M. O. Vassell, E. S. Koteles, and B Elman,
Phys. Rev. B {\bf 39}, 10~133 (1989).

\bibitem{dig} M. M. Dignam and J. E. Sipe, Phys. Rev. B {\bf 43},
4084 (1991).

\bibitem{west} T. Westgaard, Q. X. Zhao, B. O. Fimland, K. Johannessen,
and L. Johnsen, Phys. Rev. B {\bf 45}, 1784 (1992).

\bibitem{taka} Y. Takahashi, Y. Kato, S. S. Kano, S. Fukatsu, Y. Shiraki,
and R. Ito, J. Appl. Phys. {\bf 76}, 2299 (1994).

\bibitem{bay} M. Bayer, V. B. Timofeev, F. Faller, T. Gutbrod, and
A. Forchel, Phys. Rev. B {\bf 54}, 8799 (1996).

\bibitem{akim} A. V. Akimov, E. S. Moskalenko, A. L. Zhmodikov,
D. A. Mazurenko, A. A. Kaplyanskii, L. J. Challis, T. S. Cheng,
and C. T. Foxon, Fiz. Tverd. Tela {\bf 39}, 735 (1997) [Sov. Phys.
Solid State {\bf 39}, 649 (1997)].

\bibitem{bajaj} K. K. Bajaj, in {\em Properties of III-V Quantum Wells
and Superlattices}, edited by P. Bhattacharya, EMIS (Published by
INSPEC, The institution of Electrical Engineers, London, 1996),
No. 15, p. 55.

\bibitem{gap} R. P\"{a}ssler and G. Oelgart, J. Appl. Phys.
{\bf 82}, 2611 (1997).

\end{thebibliography}
\end{document}